\documentclass[preprint]{elsarticle}
\usepackage{lmodern}
\usepackage[colorlinks=True, urlcolor=Black, linkcolor=Black, citecolor=Black, allcolors=black]{hyperref}
\usepackage{amsmath, amsthm, amssymb, amsfonts}
\usepackage[nameinlink,noabbrev]{cleveref}
\usepackage{textcomp}
\usepackage{graphicx}
\usepackage{caption}
\usepackage{subcaption}
\usepackage[separate-uncertainty=true, exponent-product=\cdot]{siunitx}
\graphicspath{{Plots/}}
\usepackage{microtype}
\usepackage{lineno}

\creflabelformat{equation}{#2#1#3}

\title{Understanding the Frequency Dependence of Capacitance Measurements of Irradiated Silicon Detectors\tnoteref{t1,t2}}

\author[1]{Sven Mägdefessel\corref{cor1}}%
\author[1]{Riccardo Mori}
\author[1]{Niels Sorgenfrei}
\author[1]{Ulrich Parzefall}

\address[1]{Physikalisches Institut, Albert-Ludwigs-Universität Freiburg, Hermann-Herder-Straße 3, 79104 Freiburg, Germany}

\begin{document}

\maketitle	



\section{Introduction}
Capacitance-voltage (CV) measurements are a widely used technique in silicon detector physics. By increasing the reverse bias voltage while scanning the capacitance, it is possible to derive crucial parameters like the full depletion voltage or the doping concentration. However, this only works if the sensor is not too heavily irradiated. After irradiation, the measured CV curves show strong frequency dependencies which is not the case before irradiation and thus, the derived parameters vary over a wide range, indicating that the method is not applicable for such devices.\\
Many different models \cite{BEGUWALA, schibli} have been derived to explain frequency-dependent capacitance values but none of these were able to describe irradiated silicon detectors over the entire frequency range. In this work, an existing model for fitting CV measurements of irradiated sensors is extended. From this, it is investigated where the different frequency behaviour of the current CV analysis technique originates and it is shown how sensor parameters can be determined. Hereby, unirradiated sensors produced by a CMOS foundry \cite{DIEHL} as well as irradiated CMOS, ATLAS R0 and ATLAS R5 sensors have been CV measured over the full, technically available, frequency range and were fitted with the derived model.

\section{Setup and Measurement Procedure}
The setup used in this work is based on a common setup for CV measurements which was extended for the special needs of irradiated devices. The basic setup consists of a Keythley 237 high voltage source used to bias the sensor and simultaneously measure the leakage current. Additionally, a HP 4284 LCR meter can be (IV resp. CV configuration) AC coupled to the high voltage source to measure the capacitance without high voltage at the LCRs inputs. The sensor is placed on a chuck which is directly connected to the HV potential of the HV source and is contacted with micro positioner needles at the frontside to deliver the ground potential to the corresponding pad. The chuck consists of a top layer with embedded holes for vacuum supply and an embedded PT100 element for temperature monitoring. Underneath, peltier elements, allowing to cool the top layer to a constant temperature, are located on a cooling jig which precools the system and removes the heat of the peltier elements wherefore a Julabo chiller is used. The whole chuck stack and the micro positioners are housed in a freezer which is flushed with nitrogen to ensure a dry environment, preventing icing of the sensor while cooling down. The high voltage supply, the LCR, the chiller and the power supply of the peltier elements can be remotely controlled by the PC to acquire data or control and monitor the temperature. The LCR meter was set in parallel mode (CpRp) as this gives the best performance for low capacitances and high resistances. \\
The voltage source offers a bias voltage range up to \SI{1100}{V}, the LCR can measure capacitance, resistance or inductance in a frequency range between \SI{20}{Hz} and \SI{1}{MHz} and the cooling stack reaches temperatures, depending on the size of the sensor, down to \SI{-45}{\celsius} resp. \SI{-50}{\celsius}.\\
The measurements shown in this work were taken by ramping the voltage to a desired value, sweeping the frequency of the LCR meter across the entire offered range before measuring the current  in CV configuration. This induces more noise on the current measurement but, in contrast to ramping the sensor twice in IV and in CV configuration, systematic variations of the current are excluded. These could be caused by two effects: First, the so called hysteresis of the sensor (multiple identical runs show a systematic decrease in current) which would result in a lower current in the second run. Second, when operating the sensor close to breakdown, the current increases for some sensors with every consecutive run. A precise current measurement corresponding to every capacitance is necessary for the later analysis as both are taken into account at the same time.

\section{Theory and Model Construction}
\label{TheoModel}
The depletion width of silicon detectors features a square root dependence on the reverse bias voltage. This depletion width induces a capacitance and, when assuming a parallel plate capacitor, the following relation between the capacitance and the voltage can be deducted \cite{Sze}:

\begin{align}
	\label{C^2}
	C^{-2} = \frac{2 \epsilon_0 \epsilon_r A^2}{q_0 N_\text{eff}}V
\end{align}

\begin{figure}
	\centering
	\includegraphics[width=0.4\textwidth]{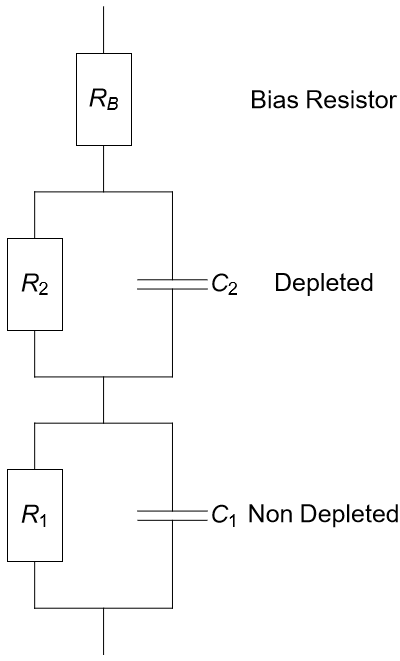}
	\caption{Equivalent circuit of an irradiated strip detector.}
	\label{EqCir}
\end{figure}

From this equation the well used method for deriving the full depletion voltage $V_\text{FD}$ and the effective doping concentration $N_\text{eff}$ can be inferred if the sensor active area $A$ is known: The inverse square of the capacitance is plotted against the voltage, a linear fit is applied to the increasing section, the flat part determines $V_\text{FD}$ and the slope is solely dependent to $N_\text{eff}$.\\
However, the physical explanation and the above shown equation both indicate that the capacitance should only be dependent on the applied voltage and not on the frequency used for the capacitance measurement. This holds true for unirradiated sensors but as the fluence exceeds a threshold of \SI{\approx\,e13}{}-\SI{e14}{\text{n}_\text{eq}}, depending on the base material, a clear dependency can be seen.\\
One reason for frequency dependence is the introduction of defects by radiation. Their description involves trapping and detrapping time constants and can thereby influence the time dependent response to a superimposed AC voltage \cite{BEGUWALA, schibli}. These time constants are in the order of \si{\micro\second} to \si{\milli\second}.  In Ref. \cite{SCHWANDT2019162418} a further explanation which introduces a frequency dependency was discussed. The basic assumption is an increasing resistivity in the non-depleted part of the sensor (which is said to be low enough to be neglected for unirradiated material). Subsequently, the resistivity of the undepleted layer and the capacitance of the depleted layer in series form an RC circuit which shields the bulk material from the oscillating voltage of the LCR. Therefore, for high frequencies, a capacitance measurement always yields the value of the sensor in full depletion. When going lower in frequency, a transition region occurs which leads to a frequency-independent regime following the CV relation given by \cref{C^2}. To further generalize our approach, a finite resistivity of the depleted bulk is introduced.\\
When measuring strip detectors (in contrast to pad diodes) another effect has to be taken into account. Each strip is biased via a bias resistor from the bias ring. This resistance also forms an RC circuit together with the bulk capacitance but, as the resistance is different to the one of the non depleted layer, this results in a further cut off frequency. Combining the beforehand described effects lead to the sensor equivalent circuit shown in \cref{EqCir} with capacitances and resistances being

\begin{align}
	\label{CapsRes}
	C_1 = \dfrac{\epsilon_0 \epsilon_r A}{d-w} \;\; \text{resp.} \;\;	C_2 = \dfrac{\epsilon_0 \epsilon_r A}{w}\\	
	R_1 = \dfrac{\rho_\text{nd} (d-w)}{A} \;\; \text{resp.} \;\; R_2 = \dfrac{\rho_\text{d} w}{A}
\end{align}

with $w$ being the depletion width, $d$ the sensor thickness, $A$ the active sensor area, $\rho_\text{nd}$ the resistivity of the non-depleted silicon bulk and $\rho_\text{d}$ the resistivity of the depleted bulk.\\

To construct a fit function from this, one has to compute the impedance of the equivalent circuit of \cref{EqCir}, equate it to the impedance of the simple RC parallel circuit assumed by the LCR meter and solve for $C_2$ and $R_2$ respectively. After this the measured capacitance and resistance can be related to the device parameters $d$, $A$, $\rho_\text{nd}$, $\rho_\text{d}$ and $R_\text{bias}$.

This model has been shown to describe the measurements quite accurately for high frequencies but not for low frequencies (already described in Ref. \cite{Li}). As mentioned before, a frequency-independent regime is expected. However, even for frequencies down to \SI{20}{Hz}, a strong dependency remains. As discovered for unirradiated measurements, it turns out that a current flowing through the sensor results in an additional capacitance contribution being proportional to $e^{-\sqrt{f}}$ which is called low-frequency (LF) increase in the following (\cref{CVincStripUnirrad}). The origin of this contribution is not completely understood but some observations give hints towards diffusion capacitance as illustrated in \cref{UnirradDevs}. It is important to mention that this term cannot be represented by either a single or multiple additional elements in the equivalent circuit as these would only introduce a broken rational function but not an exponential function.



Furthermore, it turns out that the same analytic term is also necessary to be able to fit resistance data. To account for unknown quantities in this term there are two parameters $c_1$ and $c_2$ empirically introduced in the following way 

\begin{align}
	C_\text{LF} = c_1 e^{- c_2 \sqrt{f}} \;\; \text{resp.} \;\; R_\text{LF} = c_1 e^{- c_2 \sqrt{f}}
\end{align}

\begin{figure}
	\centering
	\includegraphics[width=0.9\textwidth]{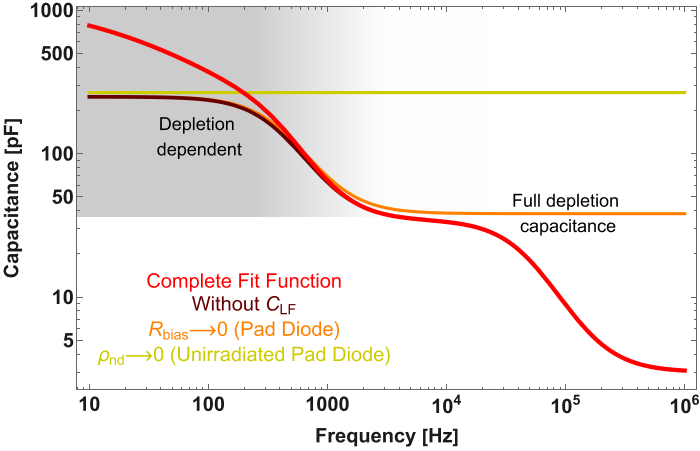}
	\includegraphics[width=0.9\textwidth]{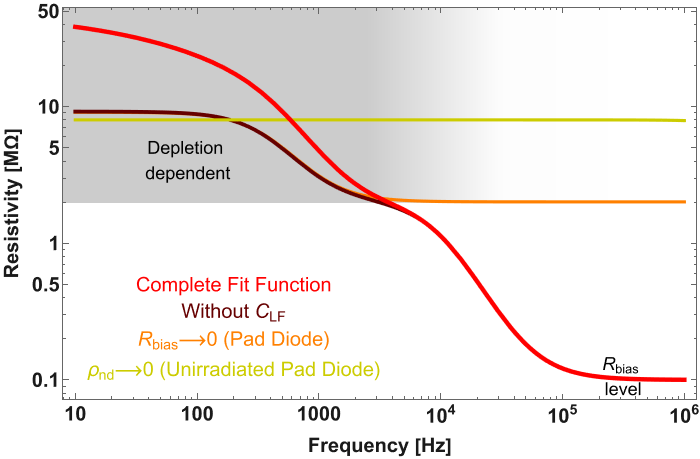}	
	\caption{Simulation of the frequency dependence of the measured capacitance and resistivity for a partially depleted sensor. The low-frequency part in grey is depletion dependent, i.e. the capacitance decreases and the resistance increases with increasing voltage.}
	\label{FitFunc}
\end{figure}

These two terms are added to the capacitance and the resistance deduced from the impedance. The fit functions are analytically computed using Wolfram Mathematica and have a complex analytic structure which makes it not possible to show here. Regarding the capacitance, it has to be noticed that not the whole sensor is biased through the bias resistors but there is a (small) area under the bias ring which is biased directly. This leads to the fact that for very high frequencies, the capacitance does not drop to exactly zero. This remaining capacitance consists of the front to backside capacitance under the bias ring as well as of the capacitance of the bias ring towards the outermost strips and the guard ring. Therefore, it is impossible to find an analytic expression for it and in this work an additional parameter $C_\text{HF}$ was introduced. In the end the fitting functions, $C(f)$ and $R(f)$, are dependent on the global parameters, $d$, $A$, $\rho_\text{nd}$, $\rho_\text{d}$, $R_\text{bias}$ and $C_\text{HF}$, which are expected to be constant for each sensor and independent of the applied voltage, as well as the individual parameters for every voltage, $w$, $c_1$ and $c_2$. The latter two should be ideally voltage-independent but for generality purposes are introduced individually for every voltage and later on their relation to physical quantities is investigated.\\
To illustrate how the frequency dependence looks like and how the above mentioned constituents contribute, a simulation of the capacitance and resistance is shown in \cref{FitFunc}.

The red line shows a complete capacitance/resistance sweep for a partially depleted irradiated sensor. In contrast to this, the brown line depicts the same function without the $c_\text{LF}$ term. Hereby, the frequency-independent level, predicted by the equivalent circuit, becomes visible. For the capacitance, this level will move further downwards with increasing depletion, resp. voltage, according to the $\sqrt{V}$ law shown in \cref{C^2}. For the resistance, it will move further up as the more resistive depletion zone grows. Looking at the resistance, the voltage-independent high-frequency limit directly yields the bias resistance divided by the number of strips. This is valid as long as it is high enough to become constant when sweeping the measurement frequency towards higher values. For lower values, the cutoff frequency, and therefore the transition, occurs at higher frequencies which might exceed the experimentally limited frequency range. In orange the function for a vanishing bias resistance (pad diode) is shown which is constant for higher frequencies, for the capacitance this value is equal to the full depletion capacitance. Finally, the line in yellow is for a vanishing resistivity of the non depleted bulk and indicates the capacitance for an unirradiated pad diode with the same area and depletion width which is completely frequency-independent.


\section{Results}


\subsection{Unirradiated Devices}
\label{UnirradDevs}
\begin{figure}
  \centering
      \includegraphics[width=0.9\textwidth]{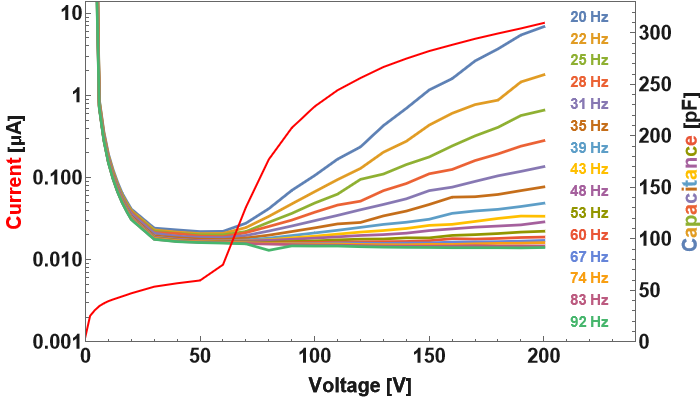}
  \caption{IV and CV measurement of an unirradiated strip detector with soft breakdown. The line in red is the current and corresponds to the axis on the left, the family of lines are the capacitance at multiple frequencies. The capacitance increase occurs simultaneously with the current increase (\SI{\approx\,60}{V}).}
  \label{CVincStripUnirrad}
  \includegraphics[width=0.9\textwidth]{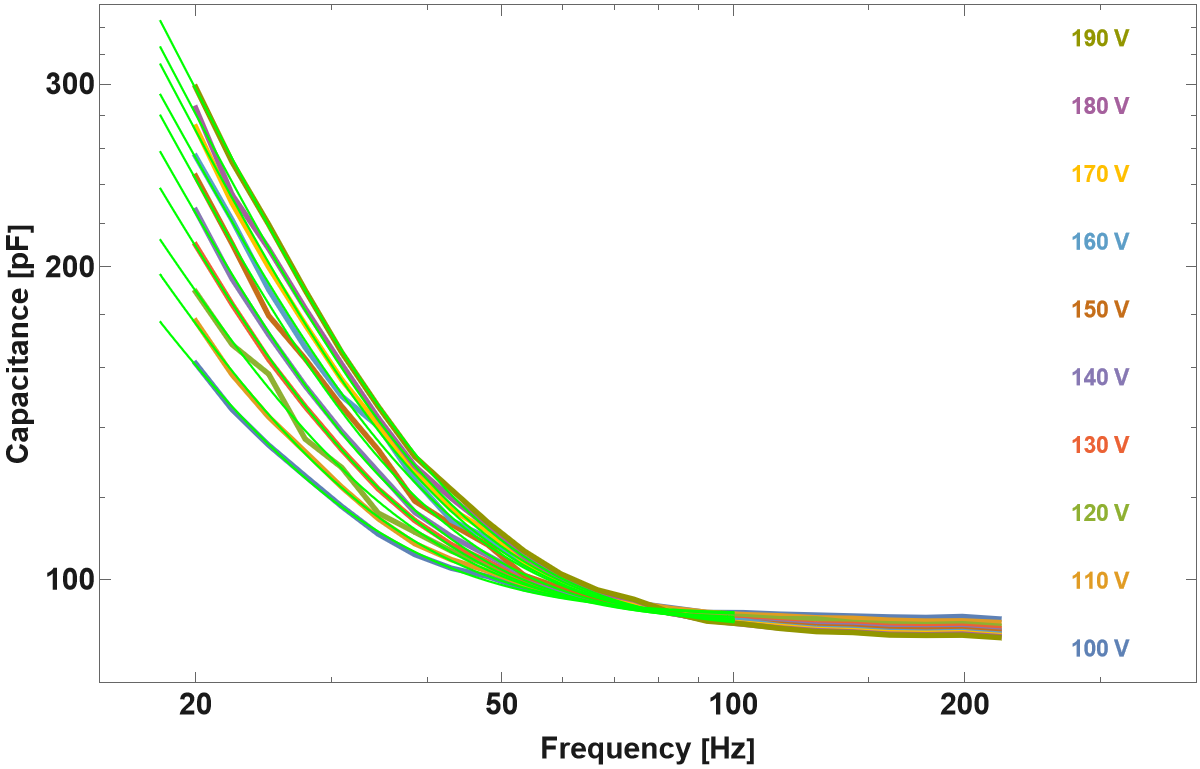}
  \caption{Low-frequency capacitance in the range of the soft breakdown. The lines in green show the fit results.}
  \label{Cincfit}
\end{figure}

As an example, the CV measurements of an unirradiated strip detector are shown in \cref{CVincStripUnirrad}. The same behaviour as shown for this detector has been observed and verified for other sensors as well.

The capacitance decreases up to voltages of around \SI{30}{V} which is the full depletion voltage of this sensor. Also, this capacitance decrease, as well as the constant level afterwards, is frequency-independent up to approximately \SI{60}{V}. Above this level, a soft breakthrough occurs which can be seen by the current increase (red line). At the same time, the capacitance starts to increase again albeit with lower magnitude for higher frequencies. This already indicates that it might be caused by the diffusion capacitance which is current driven and therefore commonly said to be neglectable in reverse bias of a diode structure.\\
To analyse the frequency dependence of the capacitance increase, the same data is plotted against the frequency as shown in \cref{Cincfit}.

For higher frequencies and voltages above full depletion (\SI{\approx\,30}{V}), the capacitance is voltage-independent (therefore the capacitance at full depletion $C_\text{FD}$ is introduced), only for frequencies below \SI{100}{Hz} an increase can be seen. It was observed that the latter obeys a $e^{-\sqrt{f}}$ law and, to test the hypothesis of the capacitance increase being influenced by the diffusion capacitance, the fit function was extended by the $\frac{\text{d}I}{\text{d}V}$ value corresponding to each voltage. In total the fit function for every voltage reads:

\begin{align}
	\label{fitfunc}
	C(V_0, f) = C_\text{FD} + c_1 \frac{\text{d}I}{\text{d}V} \bigg\rvert_{V_0} e^{- c_2 \sqrt{f}}
\end{align}

With this formula, the fits of all unirradiated sensors tested are reasonable, the values of the parameters $c_1$ and $c_2$ show no systematics among the voltages and exhibit a standard deviation of 3.8 resp. \SI{0.5}{\%}. The absence of further voltage dependencies indicates that there are no other physical quantities influencing the capacitance increase and, in particular, that the $e^{-\sqrt{f}}$ law is valid and can be taken into account for other current driven capacitance increases.


\subsection{Irradiated Devices}

\begin{figure}
	\centering
	\includegraphics[width=0.9\textwidth]{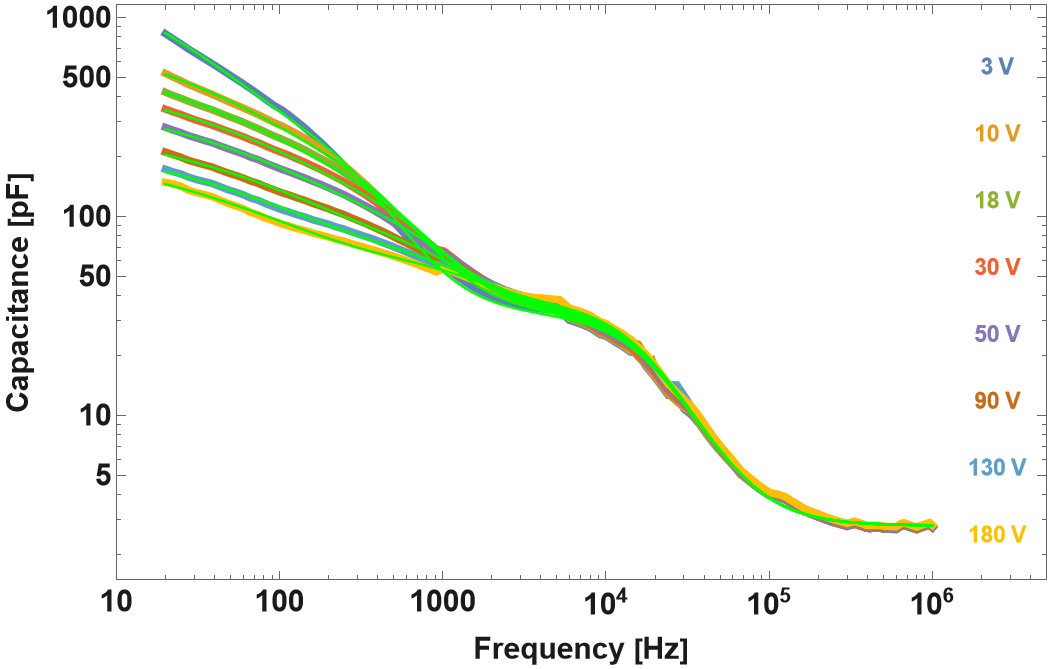}
	\caption{Full set of frequency-dependent CV measurements with according fit.}
	\label{CVFitkomplett}
\end{figure}

The derived model in \cref{TheoModel} is applied to full sets of frequency-dependent CV measurements whereby the capacitance and resistance values have been used independently. The result of a capacitance fit is shown, as an example, in \cref{CVFitkomplett}. This fit was applied to different sensors of varying temperatures and the delivered set of parameters as described in \cref{TheoModel} is analysed in the following chapters to prove the reliability of this model. 

\subsubsection{Depletion Comparison}

\begin{figure}
	\centering
	\includegraphics[width=0.9\textwidth]{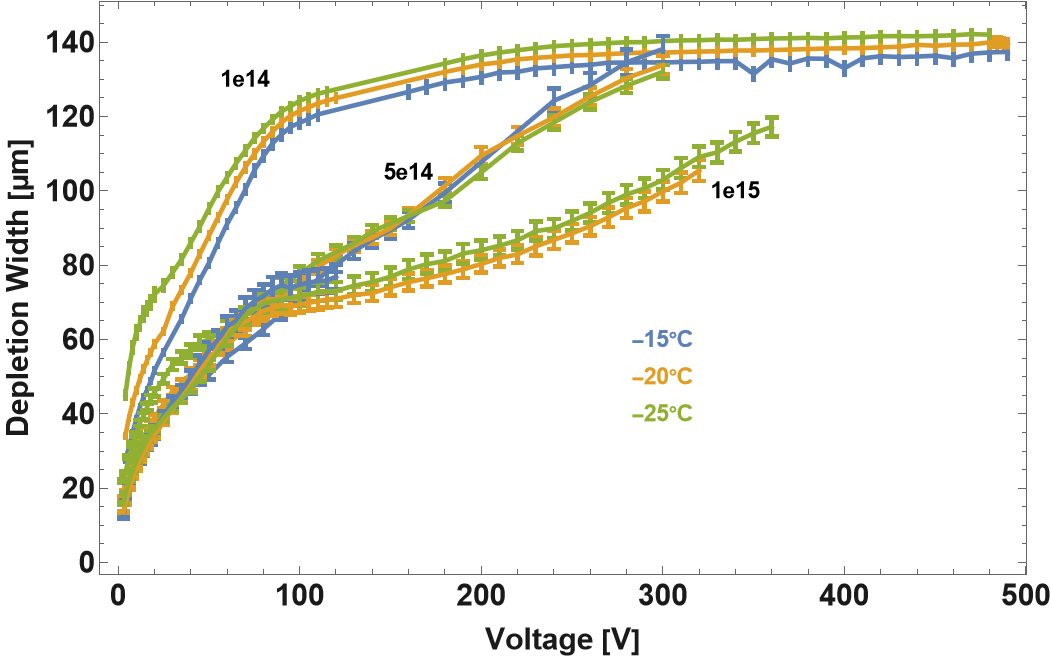}
	\includegraphics[width=0.9\textwidth]{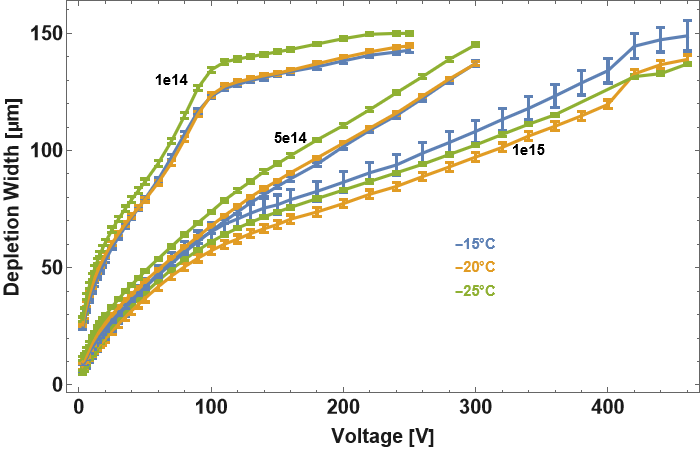}
	\caption{Depletion width of CMOS sensors irradiated to \SI{1e14}{\text{n}_\text{eq}}, \SI{5e14}{\text{n}_\text{eq}} and \SI{1e15}{\text{n}_\text{eq}} at different temperatures from capacitance data (top) and resistance data (bottom).}
	\label{DepCompCMOS}
\end{figure}

\begin{figure}
	\centering
	\includegraphics[width=0.9\textwidth]{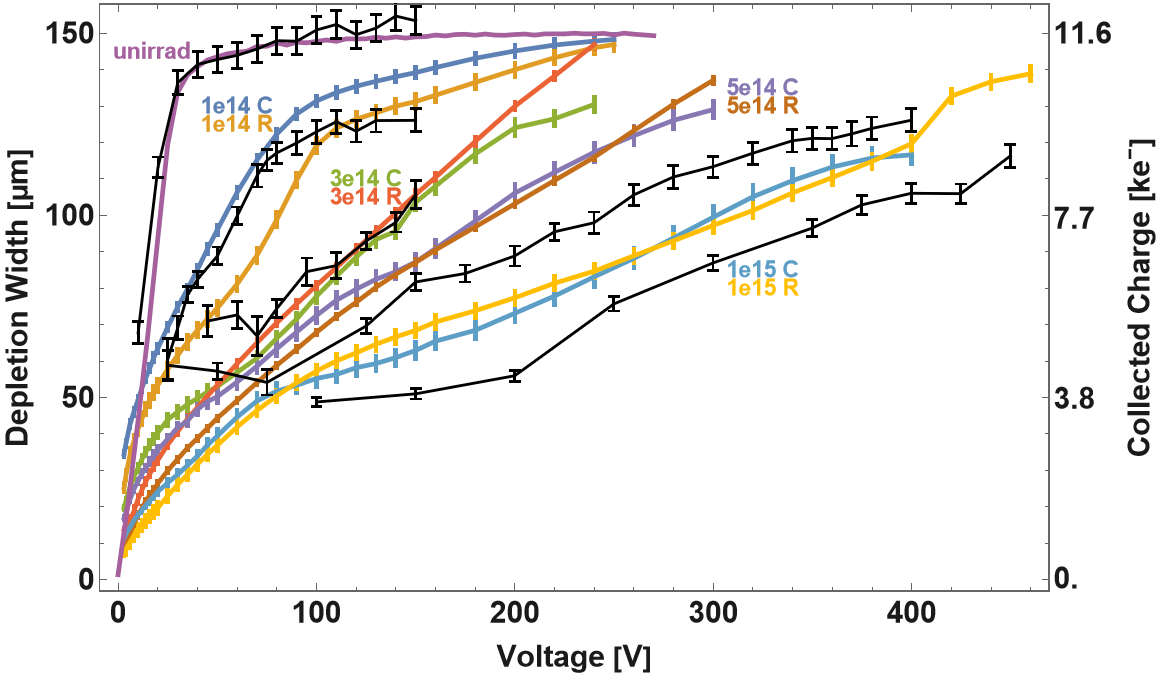}
	\caption{Depletion width of CMOS sensors irradiated to different fluences determined from capacitance and resistance data of CV measurements compared to charge collection measurements (black) of the according sensors.}
	\label{DepCompCMOSCRAlibava}
\end{figure}

First of all the depletion width of three CMOS sensors irradiated to different fluence levels is investigated. Therefore, the depletion width derived from the capacitance data as well as the width derived from the resistance data is shown at the top and bottom of \cref{DepCompCMOS} respectively. Both plots exhibit no systematic differences which indicates a good consistency of the model. It can be seen that the depletion increases with increasing voltage and the slope of the sample irradiated to \SI{1e14}{\text{n}_\text{eq}} is the steepest. At around \SI{90}{V} there is a sharp change to a constant depletion level which is slightly below the sensor thickness of \SI{150}{\micro\meter}. In contrast to expectation, the initial rise of the samples with a fluence of \SI{5e14}{\text{n}_\text{eq}} and \SI{1e15}{\text{n}_\text{eq}} is equal within the measurement uncertainties. However, above \SI{150}{V}, a clear difference in slope becomes visible. From this plot a depletion voltage of around \SI{90}{V}, \SI{300}{V} and \SI{450}{V} can be extracted. For the latter, the sensor could not be biased high enough. Because of that, the increasing depletion was extrapolated to the full depletion level of the \SI{1e14}{\text{n}_\text{eq}} measurement. Additionally, a small dependency on the temperature is visible as a constant offset towards larger depletion for lower temperatures but no significant difference in depletion voltage.\\
To verify the beforehand shown data, the CV measurements are compared to Alibava\cite{Alibava, DIEHL2} measurements of the same sensors in \cref{DepCompCMOSCRAlibava}. A fully depleted sensor of \SI{150}{\micro\meter} width is expected to collect a charge of $\SI{11.6}{\kilo e^-}$. A partially depleted sensor collects proportionally less charge due to the lower active thickness. Irradiated devices have additional charge loss channels like trapping or carrier recombination. As can be seen in \cref{DepCompCMOSCRAlibava}, the measurements for the lower fluences \SI{1e14}{\text{n}_\text{eq}} and \SI{3e14}{\text{n}_\text{eq}} show similar trends within the errors while for the higher fluences \SI{5e14}{\text{n}_\text{eq}} and \SI{1e15}{\text{n}_\text{eq}} the charge collection measurements show significantly less charge than expected from the depletion. This, however, can be explained by the additional charge loss possibilities. In conclusion, it can be stated that the depletion width determined from this CV analysis is in compliance with charge collection measurements.

\begin{figure}
	\includegraphics[width=0.9\textwidth]{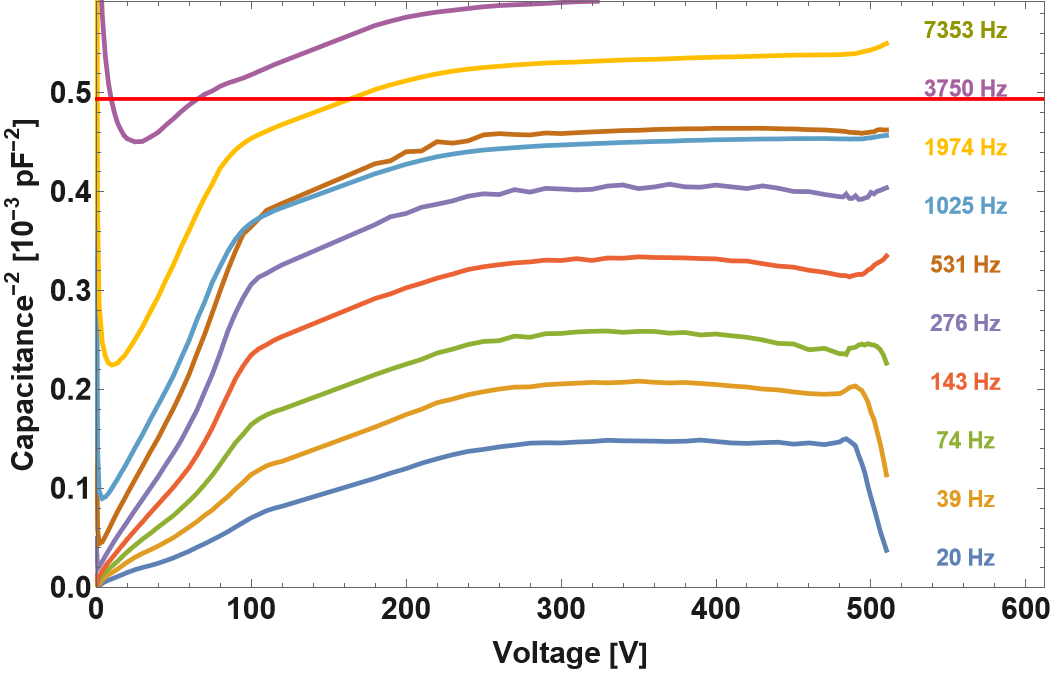}
	\caption{Inverse squared plot of the capacitance of the sensor irradiated to \SI{1e14}{\text{n}_\text{eq}} at \SI{-20}{\celsius}. The red line indicates the capacitance value of a fully depleted sensor of this geometry.}
	\label{DepComp1e14}
\end{figure}

To illustrate the problem that arises from the traditional method of deriving the full depletion capacitance or the effective doping concentration, the capacitance of the sensor irradiated to \SI{1e14}{\text{n}_\text{eq}} is plotted inverse squared in \cref{DepComp1e14}. One can see a linear increase below \SI{100}{V} but the slope is increasing with increasing frequency. Additionally, the lines for higher frequencies show an offset. This gives a lot of freedom to choose the voltage range and the frequency used for the linear fit and therefore a large spread in possible results, a trend even more strongly pronounced for higher fluences.  The reasons for the distorted $C^{-2}$ lines can be seen in \cref{FitFunc}. For low frequencies, there is an additional capacitance term, most likely induced by the diffusion capacitance. This effect reduces the $C^{-2}$ value and therefore the slope is too low. With increasing frequency this influence becomes weaker but unfortunately a further effect kicks in before it is removed completely. At a certain cutoff frequency, the irradiated bulk with its high resistivity starts to screen the depleted layer from the AC signal and therefore the capacitance starts to decrease. By choosing a frequency above the cutoff frequency it would be possible to completely avoid the influence of the low-frequency increase but the cutoff frequency itself depends on the capacitance of the depleted bulk. This is why it shifts towards higher frequencies while depleting the sensor. In contrast to the recommendation given in Ref. \cite{chilingarov}, best results are obtained by choosing the frequency where the steepest slope in the $C^{-2}$ plot is achieved (\SI{531}{Hz} in \cref{DepComp1e14}). Both beforehand mentioned effects result in a decrease of this slope but it is not possible to give a fixed number for this as the cutoff frequency depends on the temperature and sensor fluence (both shift the cutoff frequency to lower values by increasing the bulk resistivity). Nonetheless, at all frequencies an influence of either of the two effects can be observed and, when applying this method, the depletion width as well as the effective doping concentration will always be overestimated.

\subsubsection{Annealing Comparison}
\begin{figure}
	\centering
	\includegraphics[width=0.9\textwidth]{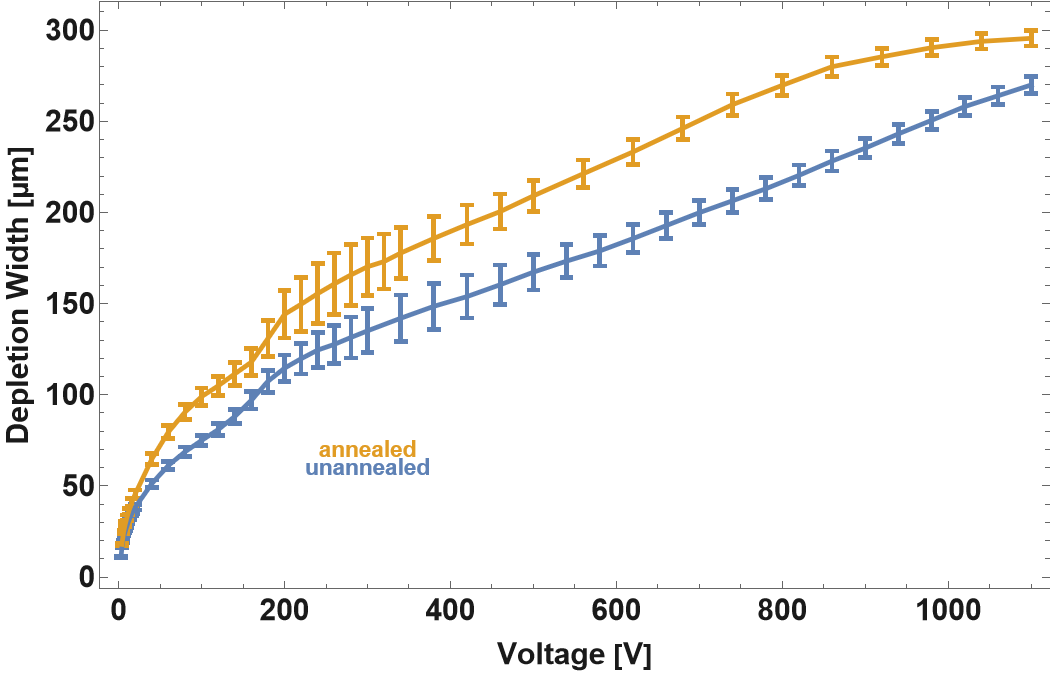}
	\caption{Depletion of an irradiated ATLAS R5 sensor before and after annealing of \SI{80}{min} at \SI{60}{\celsius}.}
	\label{AnnComp}
\end{figure}

A process used quite often to treat irradiated sensors is annealing. Hereby, the sensor is stored a fixed amount of time at elevated temperature to enable different kinds of defects to rearrange their microscopic structure and therefore also change the electrical properties. To see how large the effect of this treatment is, an irradiated ATLAS R5 sensor with a fluence of \SI{7e14}{\text{n}_\text{eq}} was annealed for \SI{80}{min} at \SI{60}{\celsius} which is expected to bring the sensor to the state of full beneficial annealing. The result can be seen in \cref{AnnComp}.

After annealing, the sensor has a higher depletion throughout the whole voltage ramp and a depletion voltage of roughly \SI{1000}{V} (active sensor thickness of \SI{305}{\micro\meter}) can be deducted while before annealing the sensor did not reach full depletion at \SI{1100}{V}. From this it can be concluded that the effective doping concentration was reduced during annealing. However, in contrast to this, in \cref{ResComp} it was found that the bulk resistivity does not change with annealing. This indicates that the mobility based correlation between the effective doping concentration and the bulk resistivity \cite{dopres} is not valid anymore for irradiated silicon.

\subsubsection{Bias Resistance Comparison}
The value of the bias resistor gained from the fits can be compared to either the values given by the sensor manufacturer or direct IV measurements between the bias pad and a single DC pad of a strip as described in Ref. \cite{TechSpec}. Before comparing the value from the fits, it has to be multiplied with the number of strips as effectively all the bias resistors are measured in parallel and therefore a much smaller value is determined. The following table shows example results of fitted bias resistance values and their corresponding comparison values.

\begin{center}
	\begin{tabular}{c||c c} 
		           &  CMOS 1e15 &  ATLAS R5 7e14 \\ [0.5ex] 
		\hline
		Fit value (\SI{-30}{\celsius}) [\si{\mega\ohm}]  & $8.2\pm0.2$ & $1.42\pm0.12$ \\ 
		\hline
		Comparison [\si{\mega\ohm}]  & $8.7\pm0.2$ & $1.5\pm0.5$\\
		\label{ResTab}
	\end{tabular}
	\captionof{table}{Bias resistance comparison for CMOS and ATLAS R5 sensors.}
	\label{biasrestable}
\end{center}

For the CMOS sensor the bias resistance is directly measured before irradiation according to Ref. \cite{TechSpec} and the value of the ATLAS R5 sensor was compared to the value given by the manufacturer Hamamatsu. It is observed that sensor annealing as well as using capacitance or resistance data for the fit does not change the result. In contrast to this, a resistance increase with decreasing temperature of $\approx -\frac{1\%}{1 K}$ is found. For irradiated sensors, the method described in Ref. \cite{TechSpec} is not applicable anymore as the irradiation causes a leakage current during depletion of the surface region. This current affects the current between the DC and the bias pad and therefore the measured resistance. To investigate whether irradiation effectively changed the bias resistance the introduced method can be used. As it is explained in \cref{TheoModel}, the most straight-forward way is to apply the highest frequency of the LCR meter as, at high frequencies, the measured resistance is equal to the bias resistance.

\subsubsection{Bulk Resistivity Comparison}
\label{ResComp}
\begin{figure}
	\centering
	\includegraphics[width=0.9\textwidth]{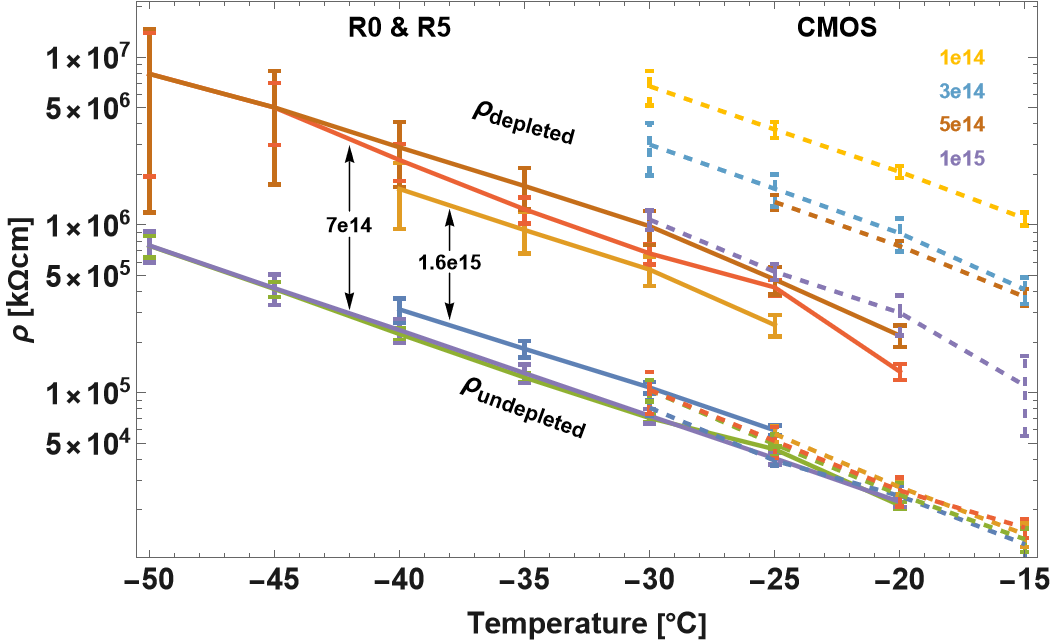}
	\caption{Resistivity of the depleted and undepleted bulk for different sensors, temperatures and fluences. For the ATLAS R5 sensors also the results prior to (red and purple) and post annealing (green and brown) are displayed.}
	\label{RhoComp}
\end{figure}

As introduced in \cref{TheoModel}, one general assumption is a change of bulk resistivity for sensors after irradiation, to be more exact, the resistivity of the depleted bulk prior to irradiation is assumed to be infinite and expected to decrease. In contrast the resistivity of the non depleted bulk is determined via the doping level and, for most sensors, given by the manufacturer. It is expected to increase as the radiation damage introduces defects to the silicon. The temperature dependent results of sensors and fluences are shown in \cref{RhoComp}. At first sight, a clear temperature dependence can be observed for the undepleted as well as for the depleted silicon. Plotting the resistivity on log scale vs. temperature, it is well in agreement with a linear trend but also the Arrhenius plot seems to be a valid approach. From an Arrhenius fit, an effective activation energy of \SI{0.60(2)}{\eV} can be determined. This is in good agreement with the position of deep lying defects within the silicon band gap. The resistivity prior to irradiation (\SI{3300}{\ohm\centi\meter}) does not show significant temperature dependence and can be determined from the effective doping concentration which itself can be derived from the full depletion voltage. The same approach applied to the full depletion voltages derived from \cref{DepCompCMOS} yields values between \SI{1.6}{\kilo\ohm\centi\meter} for \SI{1e15}{\text{n}_\text{eq}} and \SI{0.4}{\kilo\ohm\centi\meter} for \SI{1e15}{\text{n}_\text{eq}}. These values are orders of magnitudes lower than the resistivity derived by the fits. From this, it can be concluded that the resistivity increase due to irradiation is caused by deep defects and vastly stronger than the increase caused by shallow defects which only affect the effective doping concentration. 

Additionally, the influence of annealing to the resistivity was investigated. For the undepleted bulk, no change can be observed while the depleted bulk shows a resistivity increase with annealing.\\
The expectation for the fluence dependency cannot fully be confirmed. The resistivity of the non depleted bulk shows a small yet significant increase for the ATLAS sensors but no significant change within the CMOS sensors. In contrast the resistivity of the depleted bulk shows a strong decrease with increasing fluence and can be seen especially for the CMOS sensors. 

\subsubsection{Investigation of the LF Parameters}

As previously mentioned in \cref{TheoModel}, the free parameters $c_1$ and $c_2$ are introduced individually for each voltage in the fit. However, only if these values exhibit a clear signature are they likely to have a physical counterpart. Thus the voltage, temperature and fluence dependence of the free parameters of the capacitance and resistance fits are investigated in the following. Ideally, they are constant or directly dependent of another quantity.\\

\begin{figure}
	\centering
	\includegraphics[width=0.9\textwidth]{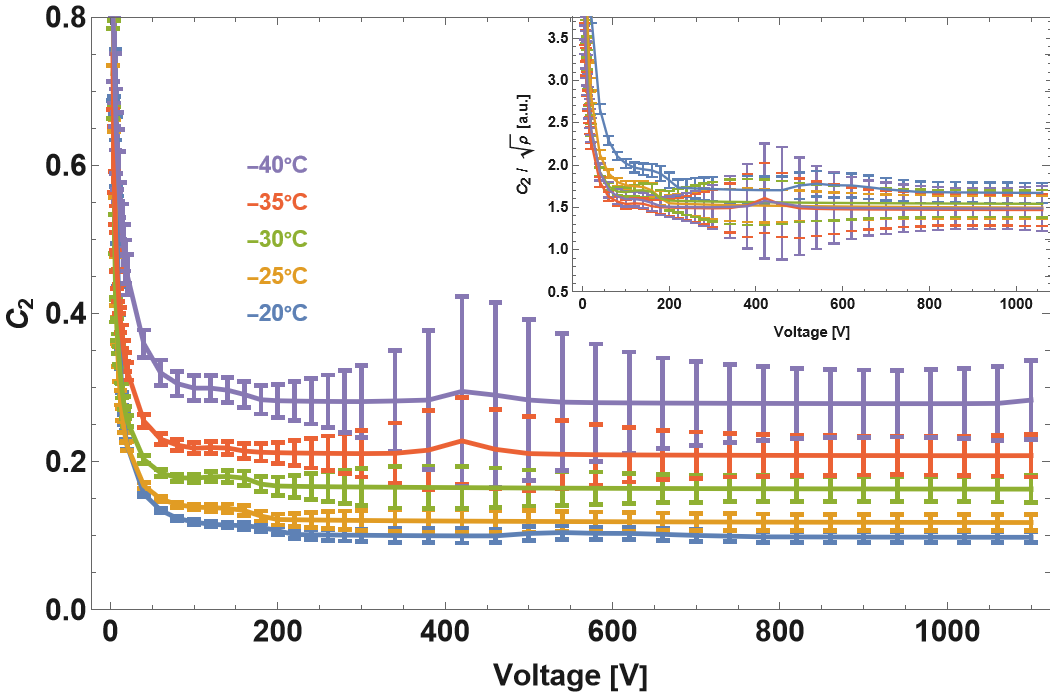}
	\caption{Voltage and temperature dependence of the free parameter $c_2$ derived from capacitance data of the ATLAS R5 sensor irradiated to \SI{7e14}{\text{n}_\text{eq}}. The inset shows the same data divided by the square root of the bulk resistivity to eliminate the temperature dependency.}
	\label{c2}
\end{figure}

At first, the voltage and temperature dependence of the parameter $c_2$ derived from capacitance data of the ATLAS R5 sensor irradiated to \SI{7e14}{\text{n}_\text{eq}} is shown in \cref{c2}. For very low voltages, values of the order of the ones obtained for the unirradiated sensors are determined but with increasing voltage, a rapid decrease followed by a saturation above \SI{100}{V} occurs. Above that threshold, there is only slight variation which is negligible compared with the errors. These show a similar trend with being small for low voltages and an increase for voltages above \SI{200}{V} and decrease again after roughly \SI{500}{V}. For the temperature dependence, a clear trend can be seen. With decreasing temperature, not only the saturation value increases, but also the errors show a steady increase. The latter one can be explained by the fact that the cutoff frequency moves to lower values with decreasing temperature and therefore the effective frequency range that can be measured decreases as the lower limit of \SI{20}{Hz} is defined by the hardware limit of the LCR. Thus a further reduction of the temperature would result in a complete inapplicability of this method. The increase with decreasing temperature is proportional to the square root of the bulk resistivity  $\rho$ as shown in the inset of \cref{c2}.\\

This general behaviour of high values decreasing rapidly and saturating was seen for all sensors irradiated to more than \SI{1e14}{\text{n}_\text{eq}}. The CMOS sensor with the lowest fluence showed the largest deviation from this behaviour which indicates that for very low fluences this general behaviour is not valid anymore. From the other sensors it can be concluded that with increasing fluence the saturation level of $c_2$ increases linearly (for example the ATLAS R0 sensor irradiated to \SI{1.6e15}{\text{n}_\text{eq}} shows roughly twice the saturation value than the ATLAS R5). A more detailed analysis of the fluence dependence would require further different fluence levels.\\

\begin{figure}
	\centering
	\includegraphics[width=0.9\textwidth]{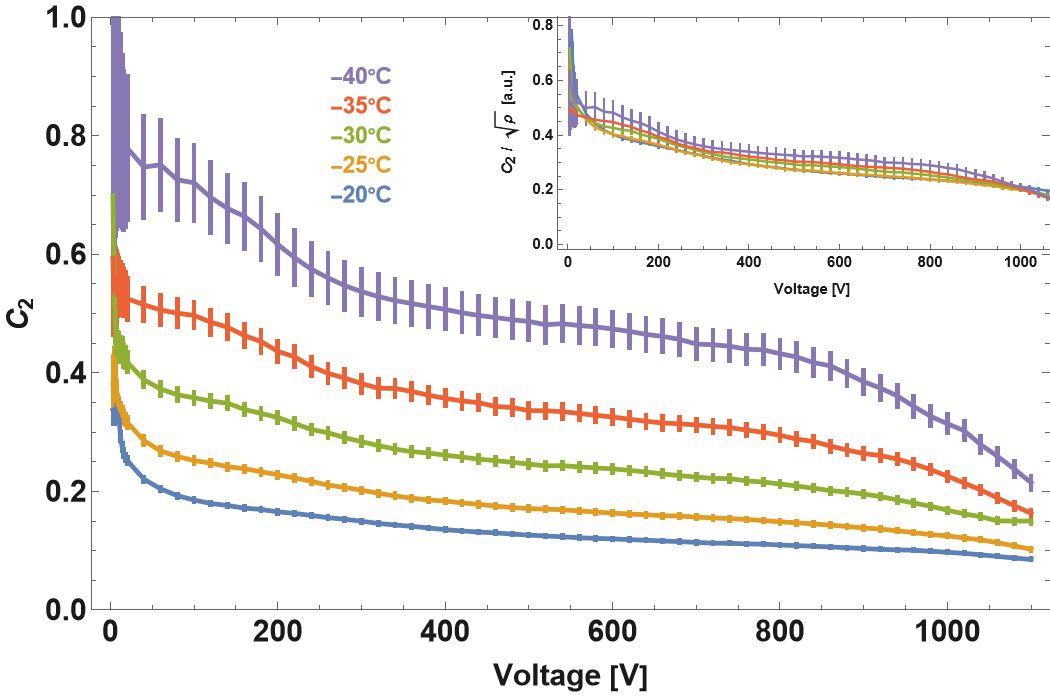}
	\caption{Voltage and temperature dependency of the free parameter $c_2$ derived from the resistance fit for the ATLAS R5 sensor irradiated to \SI{7e14}{\text{n}_\text{eq}}. The inset shows the same data divided by the square root of the bulk resistivity.}
	\label{c2res}
\end{figure}

In the same way, the $c_2$ parameter values derived from the resistance fits were analysed and are shown in \cref{c2res}. These show more of a steady decrease with increasing voltage than a stable plateau. However, the temperature trend remains the same and follows the $\sqrt{\rho}$ trend as well. The decrease in voltage cannot be related uniquely to any physical quantity alone or a combination of those.

\begin{figure}
	\centering
	\includegraphics[width=0.9\textwidth]{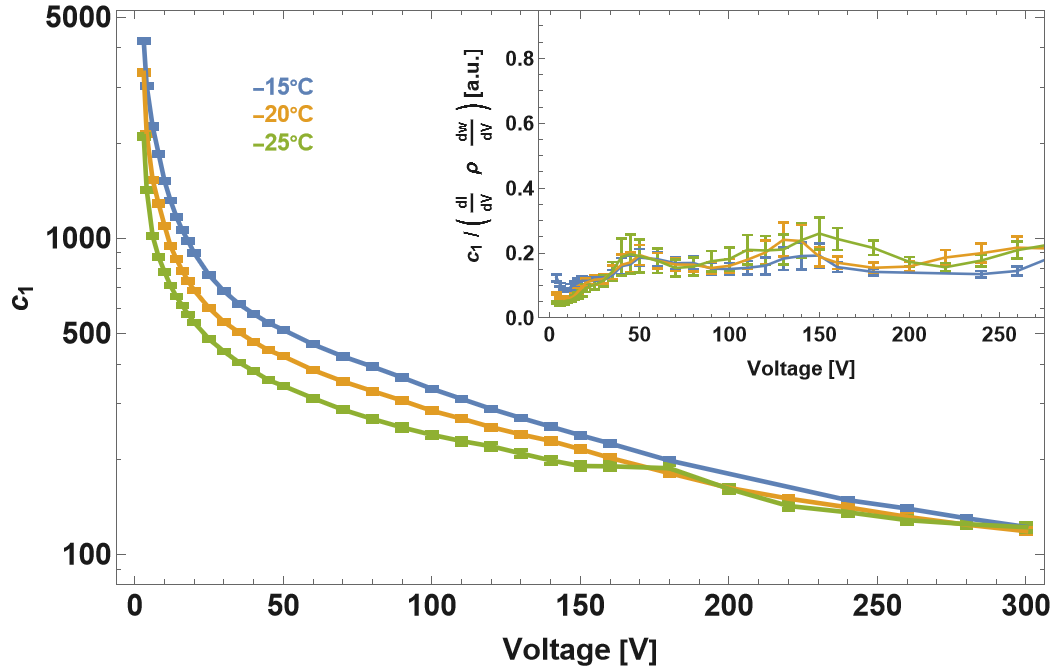}
	\caption{Voltage and temperature dependency of the free parameter $c_1$ derived from capacitance data for the CMOS sensor irradiated to \SI{5e14}{\text{n}_\text{eq}}. A decrease over multiple orders of magnitude can be seen as well as a slight temperature dependency. The inset shows the same data divided by the current increase with voltage, the bulk resistivity and the depletion increase with voltage.}
	\label{c1}
\end{figure}

Also, the dependency of the $c_1$ parameter was investigated. For all sensors, a similar trend was visible and the results of the capacitance fits of the CMOS sensor irradiated to \SI{5e14}{\text{n}_\text{eq}} are shown in \cref{c1} as an example. A strong decrease over two orders of magnitude and a temperature dependence can be seen. However, the latter one has the opposite dependency than for the $c_2$ parameter as it decreases with decreasing temperature. As already observed in \cref{UnirradDevs}, the $c_1$ parameter is dependent on $\frac{\text{d}I}{\text{d}V}$ i.e. the leakage current increase with the bias voltage. For unirradiated detectors, the consideration of this quantity is sufficient to eliminate any voltage dependencies for $c_1$. Therefore, this measure is also included for irradiated detectors now. To compensate the hereby introduced temperature dependence, the bulk resistivity (see \cref{ResComp}) has to be taken into account. Lastly, the increase in depletion width with the bias voltage $\frac{\text{d}w}{\text{d}V}$ was found to completely eliminate any voltage dependence. The resulting plot for the CMOS sensor with a fluence of \SI{5e14}{\text{n}_\text{eq}} is shown in the inset of \cref{c1}.

\begin{figure}
	\centering
	\includegraphics[width=0.9\textwidth]{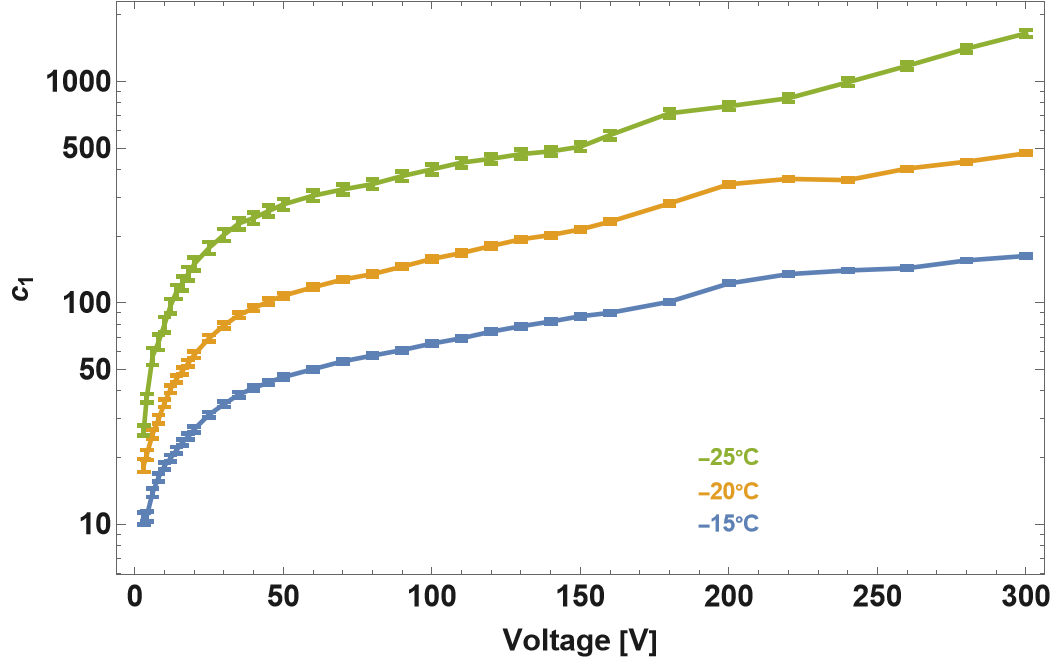}
	\caption{Voltage and temperature dependency of the free parameter $c_1$ derived from resistance data for the CMOS sensor irradiated to \SI{5e14}{\text{n}_\text{eq}}. An increase over multiple orders of magnitude can be seen and a strong temperature dependence.}
	\label{c1res}
\end{figure}

Furthermore, the dependencies of the $c_1$ parameter derived from resistance fits were under investigation (\cref{c1res}). In contrast to the capacitance fits the voltage dependency has a positive and decreasing slope. Additionally, there is a much stronger pronounced temperature dependence. The consideration of the bulk resistivity $\rho$ nearly compensates this dependency. Taking other quantities like $\frac{\text{d}I}{\text{d}V}$ or $\frac{\text{d}w}{\text{d}V}$ into account can also not fully explain the voltage behaviour.

The hypothesis of the diffusion capacitance being responsible for the LF capacitance increase can be supported by the fact that $c_1$ is dependent on the derivative of the leakage current with respect to the bias voltage. Most sources describe diffusion capacitance for forward bias and therefore carry out the derivative of the diode equation and thus conclude a linear dependence on the current. Therefore, this dependence cannot be expected for our devices in reverse bias. However, the hypothesis is contradicted by the fact that the derived frequency dependency predicts a frequency-independent low-frequency regime \cite{Sze}.

\section{Conclusion}
A novel approach for interpreting irradiated CV measurements was introduced. For this, an equivalent circuit was constructed, the corresponding impedance calculated and the capacitance a LCR instrument would measure from this was derived. In addition with an experimentally observed term, this produced a fit function able to reproduce the whole frequency range of CV measurements. As a proof of concept, multiple sensors have been measured, fitted and the gained parameter values were compared to expectations or results from other measurements and showed no evidence of model deficiencies.\\
The main new discovery is an analytical description for the current induced low-frequency capacitance behaviour. Hints towards a physical explanation are given. The question whether the diffusion capacitance is responsible for the capacitance increase cannot be finally answered.
Furthermore, it has been shown why the present technique of extracting the depletion voltage and the effective doping concentration is affected by irradiation and how to choose the least affected frequency.

\bibliographystyle{elsarticle-num} 
\bibliography{mybib}

\end{document}